\documentclass[]{emulateapj}

\def\kms{km s$^{-1}$}
\def\msun{M$_{\odot}$}

\begin{document}

\title{The Binary Fraction of Low Mass White Dwarfs}
\shorttitle{Low-Mass WDs}

\author{Justin M.\ Brown}
\affil{Franklin and Marshall College \\ 415 Harrisburg Avenue, Lancaster, PA 17604 USA}
\email{brown.justin.michael@gmail.com}

\author{Mukremin Kilic, Warren R.\ Brown, and Scott J.\ Kenyon}
\affil{Smithsonian Astrophysical Observatory \\ 60 Garden Street, Cambridge, MA 02138 USA}
\email{mkilic, wbrown, skenyon@cfa.harvard.edu}

\begin{abstract}

	We describe spectroscopic observations of 21 low-mass (\hbox{$\le$0.45 
\msun}) white dwarfs (WDs) from the Palomar-Green Survey obtained over four years.  
We use both radial velocities and infrared photometry to identify binary systems, 
and find that the fraction of single, low-mass WDs is $\leq$30\%. We discuss the 
potential formation channels for these single stars including binary mergers of 
lower-mass objects. However, binary mergers are not likely to explain the observed 
number of single low-mass WDs. Thus additional formation channels, such as enhanced 
mass loss due to winds or interactions with substellar companions, are likely.

\end{abstract}

\slugcomment{Accepted to 2011 March 10 ApJ}
%\maketitle

\section{Introduction}
\label{sec:intr}

	Around 10\% of white dwarfs (WDs) in the solar neighborhood are low-mass 
($M<$ 0.45 \msun ) helium-core objects \citep{Liebert2005}. The oldest globular 
clusters in the Galactic halo are currently producing $\approx$0.53 \msun\ WDs 
\citep{Kalirai2009}. Hence, the presence of low-mass helium-core WDs in the Galactic 
disk requires a formation mechanism other than the normal single star evolution. The 
most commonly accepted formation channel for low-mass WDs is through enhanced mass 
loss from post-main-sequence stars in interacting binary systems. Such stars can 
lose their outer envelopes without reaching the asymptotic giant branch and without 
ever igniting helium, thus ending up as helium-core WDs. The large number of 
low-mass WDs found in binary systems support this formation scenario 
\citep{Rebassa2011}. However, not all low-mass WDs are found in binary systems 
\citep{Marsh1995a, Maxted2000, Napiwotzki2004}.

	\citet{Nelemans1998} propose a scenario for the formation of single low-mass 
WDs in which binary interactions of post-main-sequence stars with close-in massive 
planets can expel the stellar envelope. The planets may or may not survive the 
common envelope evolution. In the latter case, there is no way of knowing whether 
low-mass WDs are produced through this formation channel. \citet{Nelemans1998} 
predict that the minimum companion mass required to expel the envelope of the 
progenitor 1 \msun\ star is 21 $M_{\rm J}$. Based on mid-infrared photometry, 
\citet{Kilic2010a} limit any undetected companions to $M<40 M_{\rm J}$ around eight 
single low-mass WD systems. Therefore, a well tuned common-envelope phase scenario 
involving 20-40 $M_{\rm J}$ companions is required to explain the radial velocity 
and infrared observations of the apparently single low-mass WDs. Even though this 
scenario seems unlikely, it cannot be ruled out due to the caveats in our 
understanding of the common-envelope phase evolution.

	An alternative scenario for the formation of single low-mass WDs was 
proposed by \citet{Hansen2005} to explain the color-magnitude distribution of the 
WDs in the metal-rich cluster NGC 6791.  Metal-rich stars have increased opacities 
and thus increased mass loss winds; extrapolating theoretical mass loss models to 
[Fe/H]=+0.4 suggests that mass loss can approach 0.4--0.5 \msun\ on the red giant 
branch \citep{Catelan2000}.  Thus a single low-mass WD may result from a single 
metal-rich star because of severe mass loss winds on the red giant branch. 
\citet{Kalirai2007} observe that most of the brightest WDs in NGC 6791 are indeed 
low-mass WDs, and they suggest that at least 40\% of the stars in that cluster have 
lost enough mass on the red giant branch to end up as He-core WDs. The relatively 
large fraction of low-mass WDs in NGC 6791 compared to the 10\% fraction in the 
field WD sample likely points to the dependency of stellar evolution on metallicity. 
Thus a mechanism exists to produce single low-mass WDs in metal-rich environments.

	In the mass-loss wind scenario, the formation rate of single $\sim$0.4
helium WDs is dominated by the old stellar population; intermediate-mass metal-rich
stars, even with enhanced mass loss winds, have enough mass to ignite helium in
their cores and eventually form ordinary carbon-oxygen WDs.  According to the
\citet{Reid2007} analysis of stars within 30 pc, 3\% of solar neighborhood stars
with $>$8 Gyr ages have [Fe/H]$>$+0.3.  This is comparable to the fraction of single
low-mass WDs in the PG survey WD population, which \citet{Kilic2007} estimate is
4\%.  Thus single metal-rich field stars provide another possible origin for single
low-mass WDs.

	All three scenarios mentioned above -- mass-loss due to binary companions, 
planets, and winds -- likely contribute to the population of low-mass WDs, the 
question is by how much.  The binary formation channel is generally considered to be 
the dominant mechanism, however the remaining two channels are also likely to 
contribute. Understanding the frequency of binary/single low-mass WDs is necessary 
to constrain the formation models.

	In this paper, we measure the binary frequency of the 30 low-mass WDs found
in the Palomar Green (PG) Survey \citep{Green1986}.  We provide 720 radial velocity
observations for the 21 previously unstudied low-mass WDs, and investigate the
near-infrared color excess of the whole sample.  We find that at least 70\% of the
PG WDs with $\le0.45\ M_\odot$ are binaries, and compare this observed binary
fraction and orbital period distribution to other samples of WDs and to low-mass WD
formation models.

\section{Observations and Techniques}
\label{sec:obsr}

	Our sample of thirty $\le0.45\ M_\odot$ low-mass WDs is taken from the PG
Survey, a comprehensive survey of blue stellar objects with $B\lesssim16.1$
\citep{Green1986}. The PG Survey covers 10,714 square degrees with an estimated
completeness of 84\%. \citet{Liebert2005} fit stellar atmosphere models to the 348
DA WDs in the PG survey and identify 30 WDs with mass less than 0.45 \msun.  This is
the sample of WDs studied here.  The mass and effective temperature determined by
\citet{Liebert2005} are listed in Table \ref{tab:phyp}.

\begin{deluxetable}{lcccl}
\tabletypesize{\footnotesize}
\tablecolumns{5}
\tablewidth{0pt}
\tablecaption{Physical Parameters\label{tab:phyp}}
\tablehead{
	\colhead{Object} &
	\colhead{M (\msun )} &
	\colhead{$T_{eff}$ (K)} &
	\colhead{$V$ (mag)} &
	\colhead{$J$ (mag)}
	}
\startdata
PG 0132+254 & 0.41 & 19960 & 16.12 & 16.446 $\pm$ 0.127 \\
PG 0237+242 & 0.40 & 69160 & 16.71 & 16.278 $\pm$ 0.102 \\
PG 0808+595 & 0.42 & 27330 & 16.01 & 16.608 $\pm$ 0.125 \\
PG 0834+501 & 0.40 & 60350 & 15.24 & 16.085 $\pm$ 0.104 \\
PG 0846+249 & 0.40 & 66110 & 16.71 & 17.520 $\pm$ 0.093$^*$ \\
PG 0934+338 & 0.38 & 24380 & 16.35 & 16.569 $\pm$ 0.150 \\
PG 0943+441 & 0.41 & 12820 & 13.29 & 13.643 $\pm$ 0.025 \\
PG 1022+050 & 0.44 & 11680 & 14.20 & 14.228 $\pm$ 0.129$^*$ \\
PG 1036+086 & 0.42 & 22230 & 16.26 & 16.895 $\pm$ 0.055$^*$ \\
PG 1101+364 & 0.32 & 13040 & 14.49 & 14.821 $\pm$ 0.041$^*$ \\
PG 1114+224 & 0.41 & 25860 & 16.32 & 16.970 $\pm$ 0.106$^*$ \\
PG 1202+608 & 0.40 & 58280 & 13.60 & 14.359 $\pm$ 0.032 \\
PG 1210+141 & 0.34 & 31930 & 14.71 & 15.316 $\pm$ 0.035 \\
PG 1224+309 & 0.42 & 28820 & 16.15 & 15.080 $\pm$ 0.040$^*$ \\
PG 1229$-$013 & 0.41 & 19430 & 13.79 & 14.925 $\pm$ 0.029 \\
PG 1241$-$010 & 0.40 & 23800 & 14.00 & 14.419 $\pm$ 0.043 \\
PG 1249+160 & 0.39 & 25590 & 14.62 & 15.219 $\pm$ 0.043 \\
PG 1252+378 & 0.36 & 79900 & 15.77 & 15.738 $\pm$ 0.063 \\
PG 1317+453 & 0.36 & 13320 & 14.14 & 14.317 $\pm$ 0.034 \\
PG 1320+645 & 0.44 & 27130 & 16.38 & 17.099 $\pm$ 0.079$^*$ \\
PG 1415+133 & 0.42 & 34270 & 15.37 & 14.263 $\pm$ 0.034 \\
PG 1458+172 & 0.41 & 21950 & 16.30 & 14.701 $\pm$ 0.029 \\
PG 1519+500 & 0.42 & 28730 & 16.45 & 16.950 $\pm$ 0.097$^*$ \\
PG 1554+262 & 0.45 & 21220 & 16.87 & 17.011 $\pm$ 0.083$^*$ \\
PG 1614+136 & 0.40 & 22430 & 15.24 & 15.704 $\pm$ 0.058 \\
PG 1654+637 & 0.44 & 15070 & 15.65 & 16.167 $\pm$ 0.102 \\
PG 1713+333 & 0.41 & 22120 & 14.49 & 14.900 $\pm$ 0.031 \\
PG 2226+061 & 0.44 & 15280 & 14.71 & 15.178 $\pm$ 0.044 \\
PG 2257+162 & 0.43 & 24580 & 16.14 & 15.439 $\pm$ 0.053 \\
PG 2331+290 & 0.44 & 27320 & 15.85 & 16.177 $\pm$ 0.090 \\
\enddata
	\tablecomments{ $M$, $T_{eff}$, and $V$ are from \cite{Liebert2005};
$J$ is from \cite{2Mass2003}, except for those objects ($^*$) with our own
photometry.}
\end{deluxetable}

\subsection{Photometry}
\label{sec:phot}

	We take $V$-band photometry from \citet{Liebert2005} and near-infrared
photometry from the Two Micron All Sky Survey (2MASS) \citep{2Mass2003}.  In
addition, for nine objects we obtain deeper near-infrared photometry using the
Peters Automated Infrared Imaging Telescope (PAIRITEL) -- the old 2MASS north
telescope operated with the original 2MASS camera \citep{Bloom2006}.  The photometry
of the 30 low-mass WDs is tabulated in Table \ref{tab:phyp}.

\subsection{Spectroscopy}
\label{sec:spec}

	We obtain spectroscopy for the 21 low-mass WDs with no previously
known radial velocity variability.  The other 9 WDs are known binary systems
published elsewhere \citep{Marsh1995a, Marsh1995b, Holberg1995, Orosz1999,
Morales2005, Nelemans2005b}.  To verify our data reduction and analysis procedures,
we re-observed one of the known binaries, PG 2331+290, and confirmed its 4 hr
orbital period.

	We obtained 720 spectra between October 2007 and April 2010 using the FAST
spectrograph \citep{Fabricant1998} on the Fred Lawrence Whipple Observatory 1.5m
telescope.  The spectrograph was operated with a 1.5$\arcsec$ slit and a 600 line
mm$^{-1}$ diffraction grating.  This spectrograph set-up provides a wavelength range
of 3500 to 5500 \AA\ at a spectral resolution of 1.7 \AA .  All observations were
obtained with a comparison lamp exposure following the science target exposure.

	We process the spectra using IRAF\footnote{IRAF is distributed by the
National Optical Astronomy Observatories, which are operated by the Association of
Universities for Research in Astronomy, Inc., under cooperative agreement with the
National Science Foundation.} following the guidelines in \cite{Massey1997}.  We
determine radial velocities using the cross-correlation package RVSAO
\citep{Kurtz1998} with the following procedure.  First, we measure preliminary
velocities by cross-correlating with a high signal-to-noise WD template of
known velocity.  Second, we shift each object's spectra to rest frame and sum them
together to create a high signal-to-noise template of each object.  Finally, we
cross-correlate the individual spectra with their respective template to obtain
radial velocities with the highest possible precision.  The average uncertainty of
our radial velocities is $\pm$16 \kms .

	We verify the accuracy of our velocities by measuring two night sky emission
lines, Hg 4358.335 \AA\ and 5460.750 \AA .  We find that the night sky lines have a
0 \kms\ mean and a 10 \kms\ dispersion in the data obtained in 2009 and 2010.
However, the night sky lines exhibit a larger dispersion and non-zero means up to
$\sim$10 \kms\ in the data obtained prior to 2009.  We account for this systematic
error by adding $\pm$10 \kms\ in quadrature to the radial velocity errors for
spectra obtained prior to 2009.

\section{Results}
\label{sec:resu}

	We identify binary systems among our low-mass WD sample in two ways:  by
radial velocity variability, and by infrared color excess.  We cannot detect all
binary companions, of course.  Radial velocity variability is sensitive to
companions of any type, but only for small orbital separations and non-zero
inclinations.  Infrared photometry is sensitive to companions of any orbital
separation and inclination, but only for main sequence (M dwarf) stars that produce
a significant infrared color excess.  Despite these limitations we find that at
least 70\% of our WDs have binary companions, and for these systems we estimate the
mass and nature of the companions.

\subsection{Spectroscopic Binaries}
\label{sec:fitp}

\begin{figure}
\begin{center}
\includegraphics[width=3.5in]{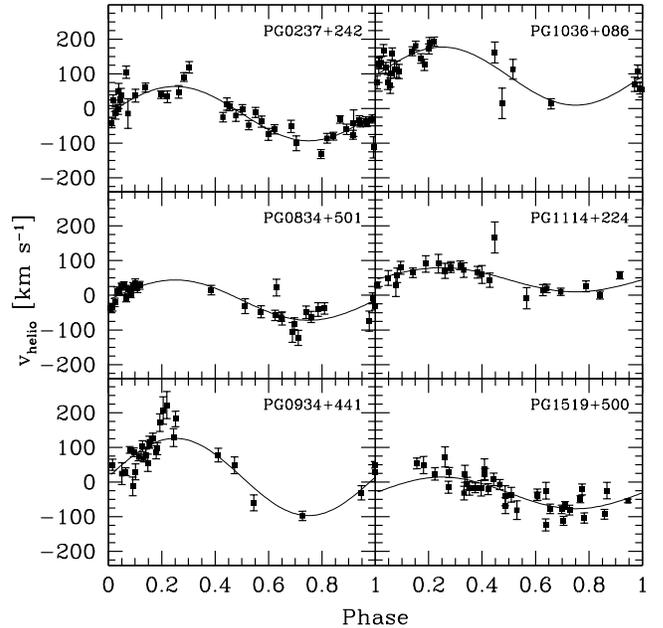}
	\caption{Radial velocities for 6 WDs with well-determined orbits, phased
to the best-fit periods (see Table \ref{tab:binp}). }
	\label{fig:phase}
\end{center}
\end{figure}

	We begin by fitting orbits to our radial velocity data, and then testing
whether the observed velocity variability is statistically significant.  We search
for orbital periods in our radial velocity data using a Lomb periodogram
\citep{Press1994}.  Period aliases are present for all of our WDs, thus we select
the period that minimizes $\chi^2$ for a circular orbital fit following
\citet{Kenyon1986}.  Figure \ref{fig:phase} (see also Figure \ref{fig:1458}) plots
the radial velocities phased to the best-fit periods for the objects with
well-determined orbits.

	We use an $F$-test to calculate the significance level at which the orbital 
fits have a smaller variance than a constant velocity fit.  For a significance 
threshold of 0.01, we find 7 WDs with significant orbital fits.  The orbital 
elements for the 7 WDs are presented in Table \ref{tab:binp}.  As discussed in 
\citet{Brown2010}, uncertainties in the orbital elements are derived from the 
covariance matrix and $\chi^2$.

	Curiously, of the WDs for which we cannot fit orbits, two objects
(PG~0846+249 and PG~1320+645) have significant velocity variability (listed in Table
\ref{tab:binp} as having ``high dispersion'').  Although we require more data to
constrain a period for these two WDs, we consider them probable binary systems.

	Table \ref{tab:binp} summarizes the orbital properties for the entire sample
of 30 low-mass WDs.  For our 7 newly discovered binaries, we present the orbital
period $P$, velocity semi-amplitude $K$, systemic velocity $\gamma$, time of
spectroscopic conjunction $T_0$, the $F$-test significance, the mass function
MF, the minimum companion mass $M_2$, and the maximum merger time $\tau$.
For the 14 WDs with no orbital fit, we present the systemic velocities and, in
place of the semi-amplitude, the standard deviation of the velocities.  Finally, for
the 9 previously known binaries, we present the published periods and mass functions
as noted in Table \ref{tab:binp}.

\subsubsection{Orbital Properties}

	Sixteen of the 30 low-mass WDs in our sample (7 from our observations and 9
from the literature) are binaries with well-measured orbits.  For these systems we
calculate the companion masses, merger times, and likelihood to produce Type Ia
supernovae ($M_{\rm tot}\geq 1.4$ \msun).

	The orbital mass function relates the observed period $P$ and velocity
semi-amplitude $K$ to the mass of the two components, $M_1$ and $M_2$, and the
inclination $i$:
	\begin{equation} \frac{P K^3}{2\pi G} = \frac{(M_2\sin{i})^3}{(M_1+M_2)^2}.
\label{eqn:mfn} \end{equation} \noindent Because the mass of the WD, $M_1$, is known
from stellar atmosphere model fits \citep{Liebert2005}, we can use the mass function
to solve for the mass of the unseen companion, $M_2\sin{i}$.  An edge-on orbit with
$i=90\degr$ gives the minimum possible companion mass and is presented in Table
\ref{tab:binp}.

	Our WD binaries are potential candidates for mergers.  A
short-period binary loses energy through gravitational wave radiation, leading to an
inward spiral of the objects until they merge \citep{Postnov2005}.  We calculate the
merger times following \citet{Landau1975},
	\begin{equation} \label{eq:merg}
\tau = \frac{(M_1 + M_2)^{\frac{1}{3}}}{M_1M_2}P^{\frac{8}{3}}\times10^{-2} \mbox{ Gyr},
	\end{equation} \noindent where the masses are in $M_\odot$, and the period
is in hours.  We use the minimum companion masses to calculate the maximum possible
merger times, presented in Table \ref{tab:binp}.

	Two of the WD+WD binaries, PG 1101+364 and PG 2331+290, will merge within a
Hubble time.  If the combined mass of the system exceeds the Chandrasekhar limit,
the merger may result in a Type Ia Supernova.  The cumulative probability function
of orbital inclination for a random stellar sample is $1-\cos{i}$
\citep{Imbert1998}.  Using this relation, we calculate the probability of the
merging binary systems becoming Type Ia Supernovae.  None of these systems have
probabilities greater than 2\% of becoming Type Ia Supernovae. Two additional
systems in our survey, PG 1224+309 and PG 1458+172 (discussed below) are WD+MS
binaries with merger times shorter than a Hubble time. These two systems will evolve
into cataclysmic binaries within several Gyr.

\begin{figure*}
\begin{center}
\includegraphics[angle=270,width=4.95in]{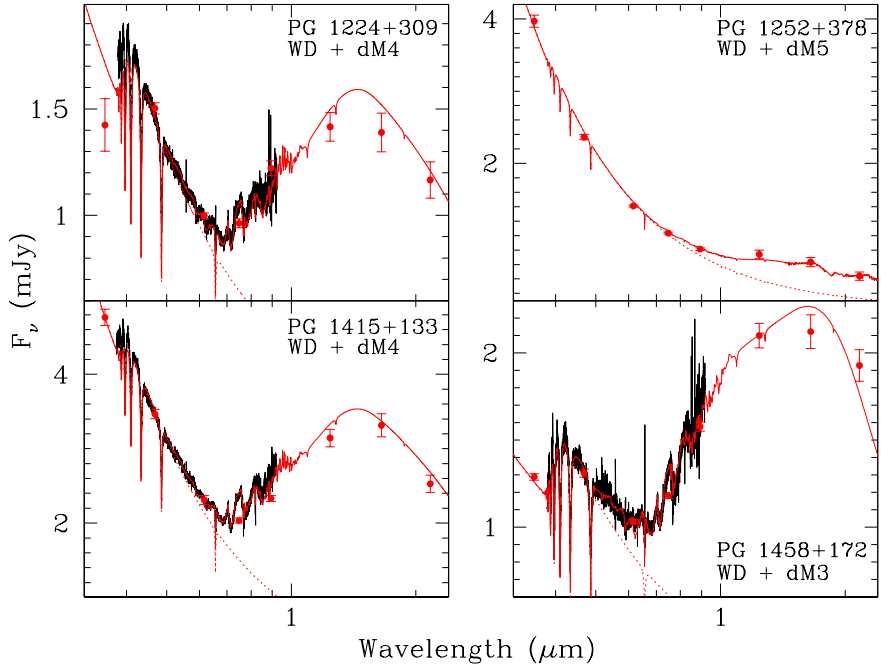}
	\caption{The spectral energy distributions for four WDs with infrared
excess.  SDSS and 2MASS photometry are plotted with red dots.  Spectroscopy is drawn
in black.  The best-fit WD (dashed red line) plus M dwarf model is drawn
with the solid red line.}
	\label{fig:sed}
\end{center}
\end{figure*}

\subsection{Photometric Binaries}
\label{sec:cole}

	We now use infrared photometry to search for binary companions to the low
mass WDs.  We begin by comparing the observed $(V-J)$ color of each WD with the
expected $(V-J)$ color from the models of \citet{Bergeron1995} for the appropriate
WD mass and effective temperature.  If there is a $(V-J)$ color excess with more than
2$\sigma$ significance, we classify the WD as having a probable (M-dwarf) companion.

	Five of the 30 WDs have significant infrared color excess, and are listed in
Table \ref{tab:cole}.  A sixth WD (PG 0237+242) with possible color excess was shown
by \citet{Kilic2010a} to have no mid-infrared signature of a companion.
\citet{Kilic2010a} also studied PG 2257+162, one of our five WDs with color excess,
and found clear evidence for a 3300 K ($\approx$ M4 spectral type) companion in the
mid-infrared {\it Spitzer} photometry.

	Figure \ref{fig:sed} plots the spectral energy distribution for the other
four WDs with color excess.  All four objects also have $ugriz$ photometry available
from the Sloan Digital Sky Survey.  In Figure \ref{fig:sed} we compare the optical
and near-infrared photometry (points) to the observed SDSS spectra (black line) and
to theoretical WD plus M-dwarf models (red lines).  All four WDs are well modeled
with main sequence M-dwarf companions with spectral types M3-M5 \citep{Pickles1998}.
Based on known M3-M5 dwarfs in eclipsing binaries, we estimate that M3, M4, and M5
dwarfs have masses of 0.28, 0.22, and 0.19 \msun, respectively \citep{Cakirli2010,
Irwin2010, Morales2009}.  The formal precision of these mass estimates is $\pm$0.05
\msun.

\subsection{Comparison of Photometry and Spectroscopy}

	Knowing the masses and the velocity amplitude (or limits to the velocity
amplitude) of the binary systems, we can place limits on the orbital periods and
inclinations using the mass function (Equation \ref{eqn:mfn}).  We separate period
and inclination by assuming, for purposes of discussion, the mean inclination angle
of a random stellar sample, $i=60\arcdeg$, and the mean orbital period observed in
our sample, $P\simeq1$ day.  Table \ref{tab:cole} summarizes the results.  The
columns in Table \ref{tab:cole} are $M_{\rm 2,phot}$, the photometric mass estimate,
$P_{{\rm For}~i=60\arcdeg}$, the orbital period given the observed velocity
amplitude and a 60$\arcdeg$ inclination, and $i_{{\rm For~P=1 day}}$, the orbital
inclination given the observed velocity amplitude and a 1 day period.

	Three systems with color excess have no detected radial velocity
variability.  The implied orbital parameters for these three systems, given the
photometric companion mass estimates, are consistent with the radial velocity
observations.  Given the 20--30 \kms\ upper limits to the velocity semi-amplitudes,
PG~1252+378, PG~1415+133, and PG~2257+162 could be binaries with either week-long
orbital periods (assuming $i=60\arcdeg$) or relatively pole-on $i<30\arcdeg$
inclinations (assuming $P$=1 day).  We would not detect any of these systems with
our radial velocity data.

\begin{deluxetable}{lccc}
\setcounter{table}{3}
\tabletypesize{\scriptsize}
\tablewidth{0pt}
\tablecolumns{4}
\tablecaption{Photometric Binaries\label{tab:cole}}
\tablehead{
	\colhead{Object} &
	\colhead{$M_{\rm 2,phot}$ (\msun )} &
	\colhead{$P_{{\rm For}~i=60\arcdeg}$ (days)} &
	\colhead{$i_{{\rm For~P=1 day}}$ ($^\circ$)}
	}
	\startdata
PG 1224+309 & 0.22 & \nodata   & \nodata     \\
PG 1252+378 & 0.19 & 5.3       & 30$\arcdeg$ \\
PG 1415+133 & 0.22 & 20.4      & 18$\arcdeg$ \\
PG 1458+172 & 0.28 & \nodata   & \nodata     \\
PG 2257+162 & 0.22 & 8.0       & 26$\arcdeg$ \\
	\enddata
\end{deluxetable}

	Two systems with color excess, PG~1224+309 and PG~1458+172, have significant
radial velocity variability. Both systems show hydrogen and magnesium emission lines
from the irradiated face of the secondary star. \citet{Orosz1999} used radial
velocity observations of PG~1224+309 to constrain the mass of the companion to
0.28 $\pm$ 0.05 \msun. This mass estimate is consistent with our photometric mass
estimate within the errors. Given the 6.2 hr period of this binary system, PG~1224+309
will evolve into a cataclysmic variable system within a Hubble time.

\begin{figure}[ht]
\begin{center}
\includegraphics[width=3in]{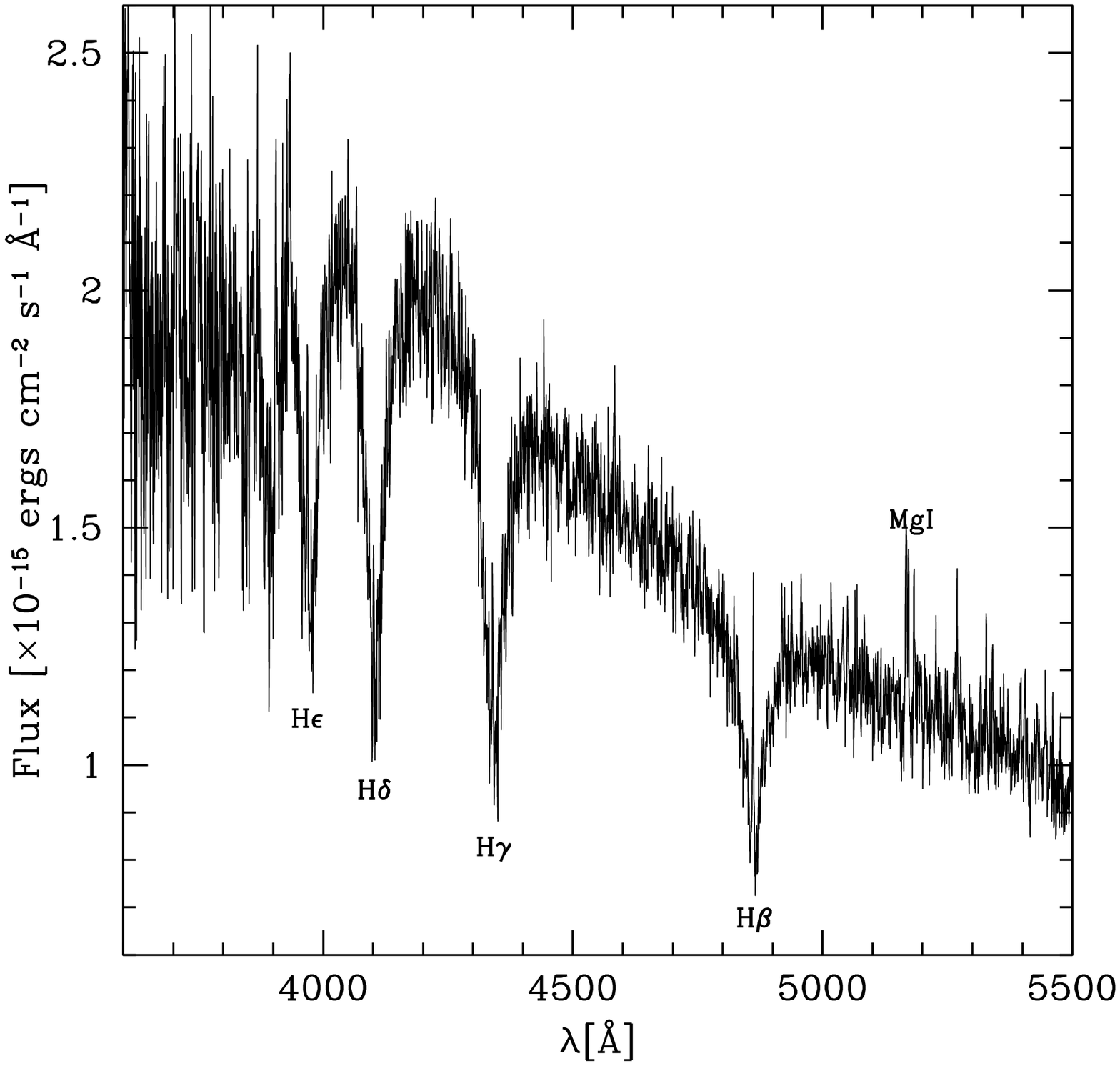}
\includegraphics[width=3in]{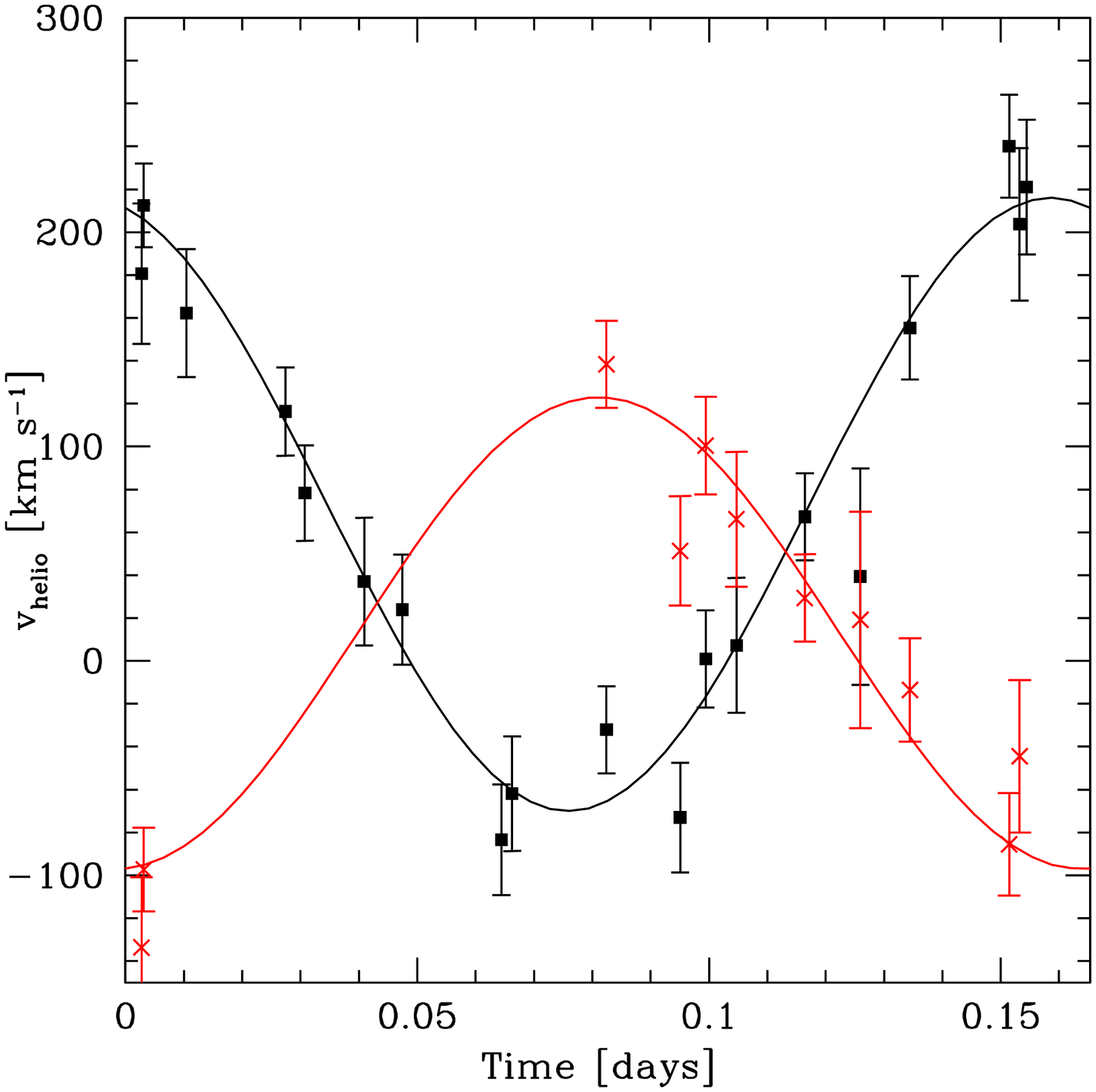}
	\caption{({\it Top}) Typical PG~1458+172 spectrum, unsmoothed, with H and Mg
lines seen in emission.  ({\it Bottom}) Observations phased to the best-fit
3.968 hr period for both absorption line velocities (squares) and emission line
velocities (x's).}
         \label{fig:1458}
\end{center}
\end{figure}

\subsection{PG 1458+172}
\label{sec:1458}

	PG~1458+172 is a new pre-cataclysmic binary discovered in our survey. The
optical spectrum of this object shows Balmer lines both in absorption and emission
and a Mg {\sc i} triplet ($\lambda$ 5167.321 \AA\ , 5172.684 \AA\ , and 5183.604
\AA ) in emission as seen in Figure \ref{fig:1458}.  Strong H$\alpha$ emission is
also evident in the SDSS spectroscopy. The radial velocity variation of the emission
lines is about 190$\arcdeg$ out of phase with the Balmer absorption lines from the
WD primary.  We use the H and Mg emission lines to fit an independent orbit to the
secondary, and list the orbital parameters in Table \ref{tab:binp} under PG~1458
(Mg).  Both the primary and secondary have best-fit periods of $0.16532 \pm 0.00033$
days.

	The ratio of the velocity amplitudes, combined with the 0.41 \msun\ WD mass
from \citep{Liebert2005}, suggests that the secondary has a mass of 0.55 \msun.
However, the center of light of the emission features is different than the center
of mass of the companion.  If the emission features come from the heated face of the
companion, for example, we would expect a shift in the velocity semi-amplitude of
the emission features \citep{Orosz1999}.  Fortunately, we detect the light of the
companion in the infrared (see Figure \ref{fig:sed}) which provides a more reliable
companion mass estimate.

	Figure \ref{fig:sed} shows that the infrared excess around PG~1458+172 is
consistent with an M3 (0.28 \msun) dwarf companion. This is significantly lower than
predicted by the mass function and the spectroscopic mass measurement for the WD.
\citet{Liebert2005}, unaware of the Balmer emission features, likely over-estimated
the WD mass, however.  Higher signal-to-noise ratio spectroscopy covering the
H$\alpha$ emission line is needed to properly characterize the orbits and masses of
the individual components in this pre-cataclysmic binary.

\subsection{Binary Fraction and Completeness}

	Of the 21 low-mass WDs we observed from the Palomar-Green Survey, seven of
these have clear periodic radial velocities, two have large enough dispersion to be
binary systems (without our being able to constrain a period), and three of those
without strong velocity variation have infrared excess, indicating a companion.
Including the nine objects from the literature that comprise the 30 low-mass WD
sample of the Palomar-Green Survey, 21 show significant evidence for a companion.
Thus the binary fraction of our $\simeq$0.4 $M_\odot$ WDs is at least 70\%.

	Some binaries must escape our detection, however.  We are not sensitive to
binary systems with intrinsically faint companions (such as brown dwarfs and WDs)
that have pole-on orientations, or large orbital separations.  For a sample of 30
binary systems with random orbital inclinations, there should be 4 pole-on systems
with $i<30\arcdeg$ for which we would not detect radial velocity variability.
Given our discovery rates, one of these hypothetical pole-on systems should be
identified by infrared photometry.  Thus there may be an additional 3 (10\%) more
binaries in our sample, undetected because they are pole-on systems.  Put another
way, 20\% -- 30\% of our low-mass WDs may exist in single systems.

	Additional observations suggest that some of the low-mass WDs are indeed 
single.  \citet{Maxted2000} have 7-15 radial velocity measurements for five of the 
possible single systems (PG 0132+254, 0808+595, 1229$-$013, 1614+136, and 2226+061) 
and do not detect significant radial velocity variations for any of these objects. 
Equally important, mid-infrared photometry by \citet{Kilic2010a} rules out the 
presence of both stellar and massive brown dwarf companions in six of the apparently 
single systems in our sample.  We conclude that possibly six (20\%), but no more 
than nine (30\%), of our 30 targets are single low-mass WDs.

\begin{figure}[ht]
\begin{center}
\includegraphics[width=3.35in]{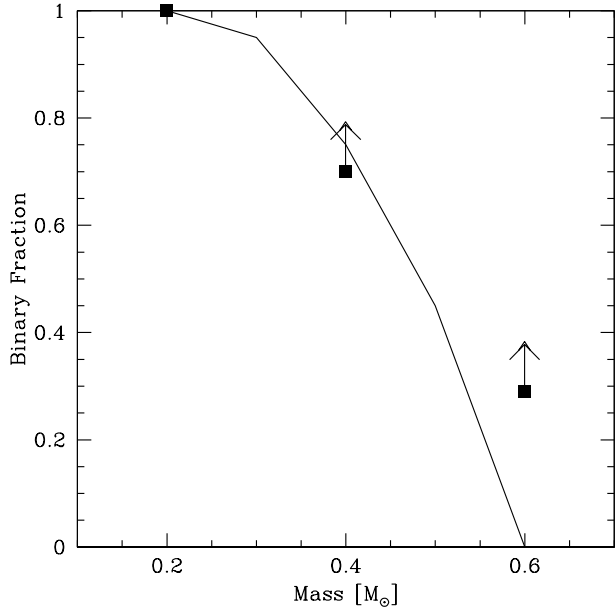}
	\caption{The observed binary fractions of WDs as a function of
mass, compared to a theoretical binary evolution model (Nelemans 2010, private
communication).  The $0.2\ M_\odot$ WDs are from \citet{Brown2010}, the
$0.4\ M_\odot$ WDs are from this paper, and the $0.6\ M_\odot$ WDs are from
\citet{Holberg2008}.
	\label{fig:frac}}
\end{center} \end{figure}

\section{Discussion}
\label{sec:disc}

\subsection{Comparison to Population Synthesis Models}

	The observed fraction of WD binary systems is related to WD mass.  WDs with
0.5-0.7 \msun\ and 0.7-1.1 \msun\ observed in the volume-limited sample of
\cite{Holberg2008} have binary fractions of at least 32\% and 6\%, respectively. The
binary fraction of extremely low-mass WDs with $\le 0.25$ \msun , on the other hand,
is 100\% \citep{Brown2010, Kilic2010c}. Hence, the binary fraction of WDs rises as
the WD mass falls.  In principle, we can use this relation between binary fraction
and WD mass to constrain binary evolution models.

	\citet{Nelemans2005a} perform population synthesis calculations for single
and double WD systems.  The predicted WD binary fraction depends on at least three
factors:  the physics of common envelope phase evolution and the initial mass
distribution and orbital period distribution of the progenitor binaries.  In these
models, for example, a portion of the single low-mass WD population forms from He+He
WD mergers \citep{Nelemans2010}.

	Figure \ref{fig:frac} compares the result of the Nelemans et al.\ population
synthesis models with the observations.  The models predict binary fractions of
100\% and 75\% for 0.2 and 0.4 \msun\ WDs, respectively. The observed binary
fractions agree remarkably well with model predictions for low-mass WDs. Given the
uncertainties in the population synthesis calculations, however, it is worth
comparing the observed binary fraction of low-mass WDs with that of single subdwarf
stars.

	The majority (90\%) of He+He WD mergers are expected to create subdwarf B
stars, the mass of which are expected to exceed the helium ignition limit of
$\approx0.45$ \msun. The birth rate of single subdwarf B stars is $2 \times
10^{-14}$ pc$^{-3}$ yr$^{-1}$ \citep{Nelemans2010}. In comparison, the formation
rate of 0.4 \msun\ WDs in the PG survey is $\sim 4 \times 10^{-14}$ pc$^{-3}$
yr$^{-1}$ \citep{Liebert2005}.  Thus the formation rate of single low-mass WDs,
given our observations, is $\sim 10^{-14}$ pc$^{-3}$ yr$^{-1}$.  This is comparable
to the birth rate of single subdwarf B stars, which means that the merger scenario
by itself cannot explain the formation rate of single low-mass WDs.

	\cite{Nelemans1998} discuss the possibility of creating single low-mass WDs
through common envelope evolution of solar mass stars with massive planets or brown
dwarfs in close ($<3$ yrs) orbits. However, the absence of stellar or massive brown
dwarf ($\geq 40 M_{\rm J}$) companions to half a dozen apparently single low-mass
WDs restricts the potential companions to a narrow mass range, where they are
massive enough to expel the stellar envelope during the common envelope phase but
also small enough to avoid detection in the mid-infrared \citep{Kilic2010a}. A well
tuned scenario involving a common envelope phase between a 1 \msun\ star and a 20-40
$M_{\rm J}$ brown dwarf at a certain orbital separation can explain the single
low-mass WDs, but it seems unlikely that this scenario will explain all single
low-mass WDs.

        \cite{Kilic2007} propose that a significant fraction of nearby field stars have
super-solar metallicity and that the single low-mass WD population can form through
enhanced mass loss from these stars.  We find up to 9 single low-mass WDs from the
348 DA WDs in the PG survey, or 2.6\%.  Similarly, \citet{Nelemans2010} find that 15
of 636 WDs (2.4\%) observed in the SPY survey are single low-mass WDs. \cite{Kilic2007}
estimate that the fraction of metal-rich stars with [Fe/H] $>$ +0.3 has been more
than 2-3\% in the past 10 Gyr. Thus the observed frequency of single low-mass WDs
is consistent with the enhanced mass-loss scenario.

\begin{figure}[ht]
\begin{center}
\includegraphics[width=3.5in]{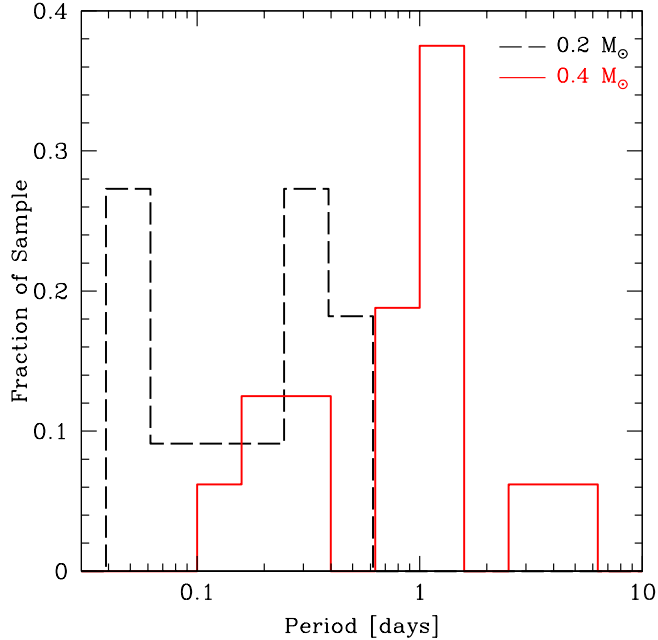}
	\caption{The period distribution of our magnitude-limited sample of 0.4
\msun\ WDs (solid histogram) compared to the magnitude-limited sample of 0.2
\msun\ WDs from \citet{Brown2010} (dashed histogram).
	\label{fig:pmhi}}
\end{center}
\end{figure}

	While binary frequency provides a useful constraint on formation scenarios,
the period distribution is fundamental to evolution of the low-mass WD binary
systems. Figure \ref{fig:pmhi} compares the observed orbital period distribution of
two complete, magnitude-limited samples:  the 0.4 \msun\ WDs studied here, and the
0.2 \msun\ WD systems studied by \citet{Brown2010}.  Interestingly, none of the 0.2
\msun\ WD systems have orbital periods longer than a day, whereas at least 40\% the
0.4 \msun\ WD systems have orbital periods longer than a day. The tightest
main-sequence binary systems will interact the earliest in their evolution;  they
will go through common envelope phases, lose their envelopes before helium ignition
in the core, and end up as He-core WDs in even tighter binary systems.  Thus finding
the lowest mass WD binaries in the most compact binary systems is consistent with
our understanding of binary evolution.

\subsection{Implications}

	We close by noting that old, metal-rich stellar populations dominate in many 
astrophysical environments, namely, the bulges and spheroids of galaxies.  The 
existence of single low-mass WDs in the field, as well as the observed stellar 
population of NGC 6791, demonstrate that old, metal-rich stars are evolving through 
at least one of the unique channels discussed here.  These evolution channels will 
affect the red giant luminosity function, blue horizontal branch morphology, and the 
integrated light of the overall stellar population \citep{Kalirai2007}. 
\citet{Han2010}, for example, argue that hot subdwarf stars formed from He-He WD 
mergers naturally explain the UV-upturn seen in the integrated light of elliptical 
galaxies.  Thus the existence of single low-mass WDs in the field, and the 
evolutionary channels they are linked to, has important implications for 
interpreting the age, metallicity, and mass-to-light ratio of metal-rich stellar 
populations throughout the Universe.

\section{Conclusions}

	We discuss radial velocity and photometric observations of 30 low-mass WDs
identified in the magnitude limited PG survey. We identify a total of 21 binary
systems and provide orbital parameters and limits on companion masses. Our sample
includes two WD+WD merger systems and two pre-cataclysmic binaries, one of which is a
new discovery.

	Nine objects in our survey do not show significant radial velocity 
variations. Six of these objects have mid-infrared photometry and none show 
mid-infrared excess from stellar or massive brown dwarf companions.  Thus the 
fraction of single, low-mass WDs is 20\%--30\%. We discuss the potential formation 
channels for these single low-mass WDs:  binary mergers of lower-mass objects, mass 
loss due to interactions with substellar companions, and mass loss due to 
super-solar metallicity winds. Binary mergers of two lower-mass WDs would most 
likely create He-burning sdB stars more massive than 0.4 \msun. The birth rates of 
single subdwarf B and low-mass WD stars are comparable. Therefore, the merger 
scenario is unlikely to explain all of the single low-mass WDs. Instead, a 
significant fraction of the single low-mass WDs may form as a result of interactions 
on the red giant branch with substellar companions or because of enhanced mass loss 
from the most metal-rich stars in the solar neighborhood.

	Our understanding of the different formation channels for single low-mass 
WDs will benefit from infrared observations that can put strong limits on potential 
substellar companions and from theoretical studies on common envelope evolution and 
the fate of short period MS + brown dwarf (or massive planet) systems. The observed 
mass and period distributions of our targets are useful for constraining binary 
evolution theory via population synthesis studies, which can probe different 
evolutionary channels for low-mass WD formation.

\section*{Acknowledgements}

	We thank G. Nelemans for providing his binary WD population synthesis model
results, Marie Machacek, Jonathan McDowell, Christine Jones, and Kara Tutunjian for
their work on the summer REU Program at the Smithsonian Astrophysical Observatory.
This work is supported in part by the National Science Foundation Research
Experiences for Undergraduates (REU) and Department of Defense Awards to Stimulate
and Support Undergraduate Research Experiences (ASSURE) programs under Grant no.\
0754568, by the Smithsonian Institution, and by NASA through the {\em Spitzer Space
Telescope} Fellowship Program, under an award from Caltech.

%\clearpage
%\bibliographystyle{/home/wbrown/lib/apj} \bibliography{bibliography}

\begin{thebibliography}{40}
\expandafter\ifx\csname natexlab\endcsname\relax\def\natexlab#1{#1}\fi

\bibitem[{{Bergeron} {et~al.}(1995){Bergeron}, {Wesemael}, \&
  {Beauchamp}}]{Bergeron1995}
{Bergeron}, P., {Wesemael}, F., \& {Beauchamp}, A. 1995, \pasp, 107, 1047

\bibitem[{{Bloom} {et~al.}(2006){Bloom}, {Starr}, {Blake}, {Skrutskie}, \&
  {Falco}}]{Bloom2006}
{Bloom}, J.~S., {Starr}, D.~L., {Blake}, C.~H., {Skrutskie}, M.~F., \& {Falco},
  E.~E. 2006, in ASP Conf.\ Ser., Vol. 351, ADASS XV, ed. {C.~Gabriel,
  C.~Arviset, D.~Ponz, \& S.~Enrique}, 751

\bibitem[{{Brown} {et~al.}(2010){Brown}, {Kilic}, {Allende Prieto}, \&
  {Kenyon}}]{Brown2010}
{Brown}, W.~R., {Kilic}, M., {Allende Prieto}, C., \& {Kenyon}, S.~J. 2010,
  \apj, 723, 1072

\bibitem[{{{\c C}ak{\i}rl{\i}} \& {Ibano{\v g}lu}(2010)}]{Cakirli2010}
{{\c C}ak{\i}rl{\i}}, {\"O}. \& {Ibano{\v g}lu}, C. 2010, \mnras, 401, 1141

\bibitem[{{Catelan}(2000)}]{Catelan2000}
{Catelan}, M. 2000, \apj, 531, 826

\bibitem[{{Cutri} {et~al.}(2003)}]{2Mass2003}
{Cutri}, R.~M. {et~al.} 2003, {2MASS All Sky Catalog of point sources.} (IPAC)

\bibitem[{{Fabricant} {et~al.}(1998){Fabricant}, {Cheimets}, {Caldwell}, \&
  {Geary}}]{Fabricant1998}
{Fabricant}, D., {Cheimets}, P., {Caldwell}, N., \& {Geary}, J. 1998, \pasp,
  110, 79

\bibitem[{{Green} {et~al.}(1986){Green}, {Schmidt}, \& {Liebert}}]{Green1986}
{Green}, R.~F., {Schmidt}, M., \& {Liebert}, J. 1986, \apjs, 61, 305

\bibitem[{{Han} {et~al.}(2010){Han}, {Podsiadlowski}, \&
  {Lynas-Gray}}]{Han2010}
{Han}, Z., {Podsiadlowski}, P., \& {Lynas-Gray}, A. 2010, \apss, 329, 41

\bibitem[{{Hansen}(2005)}]{Hansen2005}
{Hansen}, B.~M.~S. 2005, \apj, 635, 522

\bibitem[{{Holberg} {et~al.}(1995){Holberg}, {Saffer}, {Tweedy}, \&
  {Barstow}}]{Holberg1995}
{Holberg}, J.~B., {Saffer}, R.~A., {Tweedy}, R.~W., \& {Barstow}, M.~A. 1995,
  \apjl, 452, L133+

\bibitem[{{Holberg} {et~al.}(2008){Holberg}, {Sion}, {Oswalt}, {McCook},
  {Foran}, \& {Subasavage}}]{Holberg2008}
{Holberg}, J.~B., {Sion}, E.~M., {Oswalt}, T., {McCook}, G.~P., {Foran}, S., \&
  {Subasavage}, J.~P. 2008, \aj, 135, 1225

\bibitem[{{Imbert} \& {Prevot}(1998)}]{Imbert1998}
{Imbert}, M. \& {Prevot}, L. 1998, \aap, 334, L37

\bibitem[{{Irwin} {et~al.}(2010)}]{Irwin2010}
{Irwin}, J. {et~al.} 2010, \apj, 718, 1353

\bibitem[{{Kalirai} {et~al.}(2007){Kalirai}, {Bergeron}, {Hansen}, {Kelson},
  {Reitzel}, {Rich}, \& {Richer}}]{Kalirai2007}
{Kalirai}, J.~S., {Bergeron}, P., {Hansen}, B.~M.~S., {Kelson}, D.~D.,
  {Reitzel}, D.~B., {Rich}, R.~M., \& {Richer}, H.~B. 2007, \apj, 671, 748

\bibitem[{{Kalirai} {et~al.}(2009){Kalirai}, {Saul Davis}, {Richer},
  {Bergeron}, {Catelan}, {Hansen}, \& {Rich}}]{Kalirai2009}
{Kalirai}, J.~S., {Saul Davis}, D., {Richer}, H.~B., {Bergeron}, P., {Catelan},
  M., {Hansen}, B.~M.~S., \& {Rich}, R.~M. 2009, \apj, 705, 408

\bibitem[{{Kenyon} \& {Garcia}(1986)}]{Kenyon1986}
{Kenyon}, S.~J. \& {Garcia}, M.~R. 1986, \aj, 91, 125

\bibitem[{{Kilic} {et~al.}(2011){Kilic}, {Brown}, {Allende Prieto},
  {Ag{\"u}eros}, {Heinke}, \& {Kenyon}}]{Kilic2010c}
{Kilic}, M., {Brown}, W.~R., {Allende Prieto}, C., {Ag{\"u}eros}, M.~A.,
  {Heinke}, C., \& {Kenyon}, S.~J. 2011, \apj, 727, 3

\bibitem[{{Kilic} {et~al.}(2010){Kilic}, {Brown}, \& {McLeod}}]{Kilic2010a}
{Kilic}, M., {Brown}, W.~R., \& {McLeod}, B. 2010, \apj, 708, 411

\bibitem[{{Kilic} {et~al.}(2007){Kilic}, {Stanek}, \&
  {Pinsonneault}}]{Kilic2007}
{Kilic}, M., {Stanek}, K.~Z., \& {Pinsonneault}, M.~H. 2007, \apj, 671, 761

\bibitem[{{Kurtz} \& {Mink}(1998)}]{Kurtz1998}
{Kurtz}, M.~J. \& {Mink}, D.~J. 1998, \pasp, 110, 934

\bibitem[{{Landau} \& {Lifshitz}(1958)}]{Landau1975}
{Landau}, L.~D. \& {Lifshitz}, E.~M. 1958, {The classical theory of fields}
  ({Pergamon Press})

\bibitem[{{Liebert} {et~al.}(2005){Liebert}, {Bergeron}, \&
  {Holberg}}]{Liebert2005}
{Liebert}, J., {Bergeron}, P., \& {Holberg}, J.~B. 2005, \apjs, 156, 47

\bibitem[{{Marsh}(1995)}]{Marsh1995b}
{Marsh}, T.~R. 1995, \mnras, 275, L1

\bibitem[{{Marsh} {et~al.}(1995){Marsh}, {Dhillon}, \& {Duck}}]{Marsh1995a}
{Marsh}, T.~R., {Dhillon}, V.~S., \& {Duck}, S.~R. 1995, \mnras, 275, 828

\bibitem[{{Massey}(1997)}]{Massey1997}
{Massey}, P. 1997, {A User's Guide to CCD Reductions with IRAF}, {National
  Optical Astronomy Observatory}

\bibitem[{{Maxted} {et~al.}(2000){Maxted}, {Marsh}, \& {Moran}}]{Maxted2000}
{Maxted}, P.~F.~L., {Marsh}, T.~R., \& {Moran}, C.~K.~J. 2000, \mnras, 319, 305

\bibitem[{{Morales} {et~al.}(2009)}]{Morales2009}
{Morales}, J.~C. {et~al.} 2009, \apj, 691, 1400

\bibitem[{{Morales-Rueda} {et~al.}(2005){Morales-Rueda}, {Marsh}, {Maxted},
  {Nelemans}, {Karl}, {Napiwotzki}, \& {Moran}}]{Morales2005}
{Morales-Rueda}, L., {Marsh}, T.~R., {Maxted}, P.~F.~L., {Nelemans}, G.,
  {Karl}, C., {Napiwotzki}, R., \& {Moran}, C.~K.~J. 2005, \mnras, 359, 648

\bibitem[{{Napiwotzki} {et~al.}(2004){Napiwotzki}, {Karl}, {Lisker}, {Heber},
  {Christlieb}, {Reimers}, {Nelemans}, \& {Homeier}}]{Napiwotzki2004}
{Napiwotzki}, R., {Karl}, C.~A., {Lisker}, T., {Heber}, U., {Christlieb}, N.,
  {Reimers}, D., {Nelemans}, G., \& {Homeier}, D. 2004, \apss, 291, 321

\bibitem[{{Nelemans}(2010)}]{Nelemans2010}
{Nelemans}, G. 2010, \apss, 164

\bibitem[{{Nelemans} \& {Tauris}(1998)}]{Nelemans1998}
{Nelemans}, G. \& {Tauris}, T.~M. 1998, \aap, 335, L85

\bibitem[{{Nelemans} \& {Tout}(2005)}]{Nelemans2005a}
{Nelemans}, G. \& {Tout}, C.~A. 2005, \mnras, 356, 753

\bibitem[{{Nelemans} {et~al.}(2005)}]{Nelemans2005b}
{Nelemans}, G. {et~al.} 2005, \aap, 440, 1087

\bibitem[{{Orosz} {et~al.}(1999){Orosz}, {Wade}, {Harlow}, {Thorstensen},
  {Taylor}, \& {Eracleous}}]{Orosz1999}
{Orosz}, J.~A., {Wade}, R.~A., {Harlow}, J.~J.~B., {Thorstensen}, J.~R.,
  {Taylor}, C.~J., \& {Eracleous}, M. 1999, \aj, 117, 1598

\bibitem[{{Pickles}(1998)}]{Pickles1998}
{Pickles}, A.~J. 1998, \pasp, 110, 863

\bibitem[{{Postnov} \& {Yungelson}(2005)}]{Postnov2005}
{Postnov}, K. \& {Yungelson}, L. 2005, Living Rev. Rel., 9, 6

\bibitem[{Press(1994)}]{Press1994}
Press, W. 1994, Numerical Recipes in C (Cambridge: Cambridge University Press)

\bibitem[{{Rebassa-Mansergas} {et~al.}(2011){Rebassa-Mansergas}, {Nebot
  Gomez-Moran}, {Schreiber}, {Girven}, \& {Gansicke}}]{Rebassa2011}
{Rebassa-Mansergas}, A., {Nebot Gomez-Moran}, A., {Schreiber}, M., {Girven},
  J., \& {Gansicke}, B. 2011, \mnras, accepted

\bibitem[{{Reid} {et~al.}(2007){Reid}, {Turner}, {Turnbull}, {Mountain}, \&
  {Valenti}}]{Reid2007}
{Reid}, I.~N., {Turner}, E.~L., {Turnbull}, M.~C., {Mountain}, M., \&
  {Valenti}, J.~A. 2007, \apj, 665, 767

\end{thebibliography}

\begin{appendix}
%\appendix
\section{Radial Velocity Data} \label{app}

	Table \ref{tab:dat} presents our 720 radial velocity measurements for 21
low-mass WDs.  We also present 11 emission line velocity measurements, labeled
PG~1458(Mg).  The table columns are: (1) Object name, (2) heliocentric Julian date,
and (3) heliocentric radial velocity.  Table \ref{tab:dat} is available in its
entirety in machine-readable form in the online journal.  A portion of the table is
shown here for guidance regarding its form and content.

%DATA TABLE - first 5 lines
\begin{deluxetable}{lcr}
\tablecolumns{3}
\tablewidth{0pt}
\tablecaption{Radial Velocity Measurements\label{tab:dat}}
\tablehead{
 \colhead{Object} & \colhead{HJD+2450000} & \colhead{$v_{helio}$}\\
                  & (days)                & (km s$^{-1}$)
}
	\startdata
PG~0132+254 & 4385.783721 & $   38.7 \pm 14.2 $ \\
    \nodata & 4385.834625 & $   -6.3 \pm 15.5 $ \\
    \nodata & 4385.871130 & $  -18.4 \pm 13.0 $ \\
    \nodata & 4385.884278 & $   53.3 \pm 15.2 $ \\
    \nodata & 4385.909800 & $   12.3 \pm 13.3 $ \\
	\enddata
\tablecomments{Table \ref{tab:dat} is presented in its entirety in the
electronic edition of the Astrophysical Journal.  A portion is shown here for
guidance and content.  NOTE:  The complete ``data.table'' is also 
included in the arXive source file.}
\end{deluxetable}

\end{appendix}

\clearpage

%\begin{landscape}	%% BROKEN:  makes entire document landscape!
\begin{deluxetable*}{lr@{}c@{}lr@{}c@{}lr@{}c@{}lr@{}c@{}llr@{}c@{}lrrl}
\setcounter{table}{2}
\tablewidth{0pc}
\tablecaption{Binary Orbital Parameters\label{tab:binp}}
\tablehead{Object& \multicolumn{3}{c}{P (d)} & \multicolumn{3}{c}{K (km/s)}&\multicolumn{3}{c}{$\gamma$ (km/s)}&\multicolumn{3}{c}{$T_0$ (days $+$ 2450000)}&F-Test&\multicolumn{3}{c}{MF ($M_\odot$)}&$M_2$ ($M_\odot$)&$\tau$ (Gyr)&Notes }
\startdata
PG 0132$+$254 & \multicolumn{3}{c}{\dots} & $(22)$ & & & $17$ & & &\multicolumn{3}{c}{\dots} &No detectable period&\multicolumn{3}{c}{\dots} &\dots&\dots&1 \\
PG 0237$+$242 & $0.7417$ & $\pm$ & $0.0269$ & $78$ & $\pm$ & $9$ & $-14$ & $\pm$ & $4$ & $4386.6917$ & $\pm$ & $0.0154$ &$7.00e-06$& $0.0371$ & $\pm$ & $0.0134$ &$\ge 0.25$ & $\le 190$ &2 \\
PG 0808$+$595 &\multicolumn{3}{c}{\dots} &21 & $\pm$ & 7 & 28 & $\pm$ & 3 & \multicolumn{3}{c}{\dots} &$7.16e-02$&\multicolumn{3}{c}{\dots} &\dots&\dots& 1\\
PG 0834$+$501 & $1.2849$ & $\pm$ & $0.0564$ & $58$ & $\pm$ & $9$ & $-13$ & $\pm$ & $5$ & $4464.0581$ & $\pm$ & $0.0441$ &$1.47e-05$& $0.0262$ & $\pm$ & $0.0132$ &$\ge 0.22$ & $\le 920$ & \\
PG 0846$+$249 & \multicolumn{3}{c}{\dots} & $(91)$ & & & $-35$ & & &\multicolumn{3}{c}{\dots} &High dispersion&\multicolumn{3}{c}{\dots} &\dots&\dots&4 \\
PG 0934$+$338 & $1.1142$ & $\pm$ & $0.0055$ & $111$ & $\pm$ & $17$ & $14$ & $\pm$ & $10$ & $4465.7829$ & $\pm$ & $0.0205$ &$5.71e-07$& $0.1604$ & $\pm$ & $0.0759$ &$\ge 0.50$ & $\le 320$ & \\
PG 0943$+$441 & \multicolumn{3}{c}{\dots} & $(20)$ & & & $49$ & & &\multicolumn{3}{c}{\dots} &No detectable period&\multicolumn{3}{c}{\dots} &\dots&\dots&1 \\
PG 1022$+$050\tablenotemark{a} & $1.157$ & $\pm$ & $0.001$ & $75$ & $\pm$ & $1$ & $39$ & $\pm$ & $1$ & \multicolumn{3}{c}{\dots} &\dots& $0.0500$ & $\pm$ & $0.0020$ &$\ge 0.30$ & $\le 480$ & \\
PG 1036$+$086 & $1.3283$ & $\pm$ & $0.0109$ & $83$ & $\pm$ & $18$ & $93$ & $\pm$ & $11$ & $4474.7431$ & $\pm$ & $0.0499$ &$1.55e-04$& $0.0815$ & $\pm$ & $0.0525$ &$\ge 0.37$ & $\le 610$ & \\
PG 1101$+$364\tablenotemark{b} & $0.145$ & $\pm$ & $0.001$ & $71$ & $\pm$ & $2$ & $39$ & $\pm$ & $1$ & \multicolumn{3}{c}{\dots} &\dots& \multicolumn{3}{c}{\dots} &$0.37$ & $2$ &4 \\
PG 1114$+$224 & $0.3198$ & $\pm$ & $0.0147$ & $34$ & $\pm$ & $7$ & $44$ & $\pm$ & $3$ & $4476.7679$ & $\pm$ & $0.0113$ & $1.46e-03$& $0.0014$ & $\pm$ & $0.0009$ &$\ge 0.07$ & $\le 64$ & \\
PG 1202$+$608\tablenotemark{c} & $1.493$ & $\pm$ & $0.001$ & $77$ & $\pm$ & $8$ & $0$ & $\pm$ & $6$ & \multicolumn{3}{c}{\dots} &\dots& $0.0720$ & $\pm$ & $0.0200$ &$\ge 0.34$ & $\le 93$ & \\
PG 1210$+$141\tablenotemark{d} & $0.642$ & $\pm$ & $0.001$ & $131$ & $\pm$ & $3$ & $15$ & $\pm$ & $2$ & \multicolumn{3}{c}{\dots} &\dots& $0.1490$ & $\pm$ & $0.0100$ &$\ge 0.46$ & $\le 90$ & \\
PG 1224$+$309\tablenotemark{e} & $0.260$ & $\pm$ & $0.001$ & $112$ & $\pm$ & $14$ & $-4$ & $\pm$ & $12$ & \multicolumn{3}{c}{\dots} &\dots& $0.0380$ & $\pm$ & $0.0040$ & 0.28 & $\le 10$ &2 \\
PG 1229$-$013 &\multicolumn{3}{c}{\dots} &22 & $\pm$ & 7 & 22 & $\pm$ & 4 & \multicolumn{3}{c}{\dots} &$2.48e-02$&\multicolumn{3}{c}{\dots} &\dots&\dots& 1 \\
PG 1241$-$010\tablenotemark{f} & $3.347$ & $\pm$ & $0.001$ & $68$ & $\pm$ & $1$ & $16$ & $\pm$ & $1$ & \multicolumn{3}{c}{\dots} &\dots& $0.1110$ & $\pm$ & $0.0050$ &$\ge 0.42$ & $\le 6700$ & \\
PG 1249$+$160 & \multicolumn{3}{c}{\dots} & $(16)$ & & & $1$ & & &\multicolumn{3}{c}{\dots} &No detectable period &\multicolumn{3}{c}{\dots} &\dots&\dots&1 \\
PG 1252$+$378 &\multicolumn{3}{c}{\dots} &30 & $\pm$ & 8 & 26 & $\pm$ & 4 & \multicolumn{3}{c}{\dots} & $3.00e-02$&\multicolumn{3}{c}{\dots} &\dots&\dots& 2 \\
PG 1317$+$453\tablenotemark{f} & $4.872$ & $\pm$ & $0.001$ & $64$ & $\pm$ & $1$ & $-24$ & $\pm$ & $1$ & \multicolumn{3}{c}{\dots} & \dots& $0.1320$ & $\pm$ & $0.0040$ &$\ge 0.44$ & $\le 19000$ & \\
PG 1320$+$645 & \multicolumn{3}{c}{\dots} & $(110)$ & & & $-16$ & & &\multicolumn{3}{c}{\dots} &High dispersion&\multicolumn{3}{c}{\dots} &\dots&\dots&3 \\
PG 1415$+$133 &\multicolumn{3}{c}{\dots} &20 & $\pm$ & 8 & 62 & $\pm$ & 4 & \multicolumn{3}{c}{\dots} &$7.97e-02$&\multicolumn{3}{c}{\dots} &\dots&\dots& 2 \\
PG 1458$+$172 & $0.1653$ & $\pm$ & $0.0003$ & $143$ & $\pm$ & $8$ & $73$ & $\pm$ & $12$ & $4911.8550$ & $\pm$ & $0.0002$ &$1.08e-04$& 0.1805 & & & 0.55 & $\le 1.7$ &2,3,4 \\
PG 1458 (Mg) & $0.1653$ & $\pm$ & $0.0003$ & $106$ & $\pm$ & $14$ & $13$ & $\pm$ & $22$ & $4913.7613$ & $\pm$ & $0.0003$ &$6.74e-03$& (0.0386) & & & (0.21) & ($\le 5.3$) & \\
PG 1519$+$500 & $0.8603$ & $\pm$ & $0.0200$ & $45$ & $\pm$ & $9$ & $-31$ & $\pm$ & $4$ & $4270.1663$ & $\pm$ & $0.0636$ & $8.90e-04$& $0.0085$ & $\pm$ & $0.0051$ &$\ge 0.14$ & $\le 460$ & \\
PG 1554$+$262 &\multicolumn{3}{c}{\dots} &39 & $\pm$ & 10 & 0 & $\pm$ & 5 & \multicolumn{3}{c}{\dots} &$1.58e-02$&\multicolumn{3}{c}{\dots} &\dots&\dots& 1\\
PG 1614$+$136 &\multicolumn{3}{c}{\dots} &25 & $\pm$ & 6 & -7 & $\pm$ & 3 & \multicolumn{3}{c}{\dots} &$1.44e-02$&\multicolumn{3}{c}{\dots} &\dots&\dots& 1\\
PG 1654$+$637 &\multicolumn{3}{c}{\dots} &25 & $\pm$ & 9 & 29 & $\pm$ & 4 & \multicolumn{3}{c}{\dots} &$4.63e-02$&\multicolumn{3}{c}{\dots} &\dots&\dots& 1\\
PG 1713$+$333\tablenotemark{f} & $1.127$ & $\pm$ & $0.001$ & $56$ & $\pm$ & $1$ & $56$ & $\pm$ & $1$ & \multicolumn{3}{c}{\dots} &\dots& $0.0200$ & $\pm$ & $0.0010$ &$\ge 0.19$ & $\le 700$ & \\
PG 2226$+$061 & \multicolumn{3}{c}{\dots} & $(23)$ & & & $31$ & & &\multicolumn{3}{c}{\dots} &No detectable period&\multicolumn{3}{c}{\dots} &\dots&\dots&1 \\
PG 2257$+$162 &\multicolumn{3}{c}{\dots} &27 & $\pm$ & 7 & 25 & $\pm$ & 3 & \multicolumn{3}{c}{\dots} &$2.30e-02$&\multicolumn{3}{c}{\dots} &\dots&\dots& 2 \\
PG 2331$+$290\tablenotemark{f} & $0.166$ & $\pm$ & $0.001$ & $156$ & $\pm$ & $3$ & $-11$ & $\pm$ & $3$ & \multicolumn{3}{c}{\dots} &\dots& $0.0660$ & $\pm$ & $0.0040$ &$\ge 0.34$ & $\le 2$ & \\
\enddata
	\tablerefs{(a) \cite{Morales2005}, (b) \cite{Marsh1995b}, (c)
\cite{Holberg1995}, (d) \cite{Nelemans2005b}, (e) \cite{Orosz1999}, (f)
\cite{Marsh1995a}}
	\tablecomments{(1) Probable single WD. (2) Infrared color excess, M dwarf
companion. (3) Velocity variable, unconstrained orbit. (4) Double-lined
spectroscopic binary.}

\end{deluxetable*}
\clearpage
%\end{landscape}

\end{document}